\documentclass[prl,twocolumn,superscriptaddress,showpacs]{revtex4}
\usepackage{graphicx,amsmath,amssymb,bm}

\newcommand{\bra}[1]{\left\langle #1 \right|}
\newcommand{\ket}[1]{\left| #1 \right\rangle}

\renewcommand{\epsilon}{\varepsilon}
\newcommand{\be}{\begin{equation}}
\newcommand{\ee}{\end{equation}}

\newcommand{\fmi}{\, \text{fm}^{-1}}
\newcommand{\mev}{\, \text{MeV}}

\begin{document}

\title{In-Medium Similarity Renormalization Group for Open-Shell Nuclei}

\author{K.\ Tsukiyama}
\email[E-mail:~]{tsuki@cns.s.u-tokyo.ac.jp}
\affiliation{Center for Nuclear Study, Graduate School of Science,
University of Tokyo, Hongo, Tokyo, 113-0033, Japan}

\author{S.\ K.\ Bogner}
\email[E-mail:~]{bogner@nscl.msu.edu}
\affiliation{National Superconducting Cyclotron Laboratory
and Department of Physics and Astronomy, Michigan State University,
East Lansing, MI 48844, USA}

\author{A.\ Schwenk}
\email[E-mail:~]{schwenk@physik.tu-darmstadt.de}
\affiliation{ExtreMe Matter Institute EMMI, GSI Helmholtzzentrum f\"ur
Schwerionenforschung GmbH, 64291 Darmstadt, Germany}
\affiliation{Institut f\"ur Kernphysik, Technische Universit\"at
Darmstadt, 64289 Darmstadt, Germany}

\begin{abstract}
We present a nonperturbative derivation of effective valence-shell
Hamiltonians in the framework of the recently developed in-medium
similarity renormalization group (IM-SRG). As a first application,
we calculate the spectra of $p$- and $sd$-shell nuclei, $^6$Li and
$^{18}$O, based on evolved chiral nucleon-nucleon interactions. For
$^6$Li, the spectrum is in very good agreement with ab-initio
results. For $^{18}$O, the IM-SRG provides a new method for the
shell model to systematically go beyond effective interaction
techniques based on diagrammatic expansions.
\end{abstract}


\maketitle

{\it Introduction.--} Advances in ab-initio methods for nuclear
structure combined with nuclear forces based on chiral effective field
theory (EFT) have lead to many exciting developments for light nuclei
and medium-mass nuclei around closed-shell configurations (see, e.g.,
Refs.~\cite{abinitio}). For open-shell systems with many valence
nucleons, however, the shell model remains the most successful
approach to understand and predict nuclear structure, including the
evolution of shell structure with changing neutron and proton numbers,
properties of ground and excited states, and electroweak
transitions~\cite{SM}. Moreover, the shell model has recently revealed
new insights to the impact of long-range tensor~\cite{tensor} and
three-nucleon forces~\cite{3N} in neutron-rich nuclei. These are
dominated by pion exchanges, which provides a link between nuclear
structure and developments in chiral EFT interactions. Despite the
many successes of the shell model, the microscopic derivation of
effective interactions and operators among valence nucleons from
nuclear forces is still largely based on perturbative approaches where
the convergence remains an open problem.

In this Letter, we present a new nonperturbative derivation of
effective valence-shell Hamiltonians in the framework of the in-medium
similarity renormalization group (IM-SRG), which we recently developed
for closed-shell nuclei~\cite{Koshiroh,KoshirohPhD}. The IM-SRG is
based on a renormalization group evolution that decouples degrees of
freedom that are not relevant for the problem of interest. We show how
the IM-SRG can be generalized to open-shell systems away from
doubly-magic nuclei. For $^6$Li, we present first results for the
ground-state energy and spectrum in very good agreement with
ab-initio methods. We then discuss for $^{18}$O how the IM-SRG goes
significantly beyond effective interaction techniques based on
diagrammatic expansions~\cite{Morten}, opening up a promising new
method to connect nuclear forces and the shell model.

{\it IM-SRG and generator choices.--} The IM-SRG starts from a
Hamiltonian $H$ that is normal ordered with respect to a
finite-density reference state $\ket{\Phi}$ (e.g., the Hartree-Fock
ground state):
\begin{equation}
H = E_0 + \sum_{ij} f_{ij} \, \{a_i^\dagger a_j\}
+ \frac{1}{2!^2} \sum_{ijkl} \Gamma_{ijkl} \, \{a_i^\dagger a_j^\dagger
a_l a_k\} \,,
\end{equation}
where the normal-ordered strings of creation and annihilation
operators obey $\langle\Phi|\{a^{\dagger}_i \cdots a_j\}|\Phi\rangle =
0$. We include normal-ordered 0-, 1-, and 2-body operators, $E_0$,
$f$, and $\Gamma$, which approximately include induced 3- and
higher-body interactions, and solve the IM-SRG flow equations to
obtain the evolved Hamiltonian $H(s)$~\cite{Koshiroh}. We refer to
this truncation as IM-SRG(2) since we keep up to normal-ordered 2-body
operators. The evolution is equivalent to a series of unitary
transformations that are designed to evolve $H(s)$ as $s \to \infty$
to an appropriately defined ``diagonal'' part 
$H^{\rm d}(s)$~\cite{SRG,PPNP}:
\begin{equation}
H(s) = U(s) H U^\dagger(s) \equiv H^{\rm d}(s) + H^{\rm od}(s)
\rightarrow H^{\rm d}(\infty) \,.
\end{equation}
The unitary transformation $U(s)$ is determined by the generator
$\eta(s) \equiv [dU(s)/ds] \, U^{\dagger}(s)$, which is constructed
from the diagonal part,
\begin{equation}
\eta(s) = [H^{\rm d}(s),H(s)] = [H^{\rm d}(s),H^{\rm od}(s)] \,,
\end{equation}
and guarantees that the ``off-diagonal'' coupling $H^{\rm od}$ is  
driven to zero with increasing $s$.

For the ground state of closed-shell nuclei, one eliminates all terms
that couple the reference state $\ket{\Phi_c}$ to the rest of the
Hilbert space. This is achieved when the matrix elements between
$\ket{\Phi_c}$ and all $n$-particle--$n$-hole states vanish,
$\bra{npnh} H(\infty) \ket{\Phi_c}=0$. Therefore, one takes $H_c^{\rm
od}$ to be composed of all 1- and 2-body operators that connect hole
($h$) with particle ($p$) states so that $\{H_c^{\rm od}\} = \{f_{ph},
\Gamma_{pp'hh'}\}$ plus hermitian conjugates, as was demonstrated in
Ref.~\cite{Koshiroh}.

\begin{figure}[t]
\begin{center}
\includegraphics[width=0.425\textwidth,clip=]{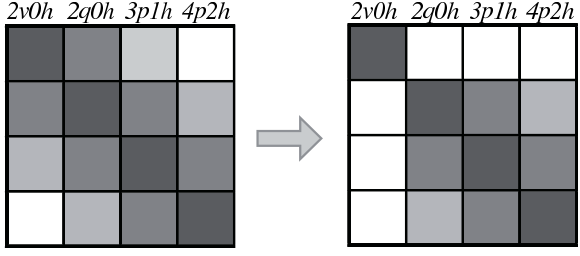}
\end{center}
\caption{Schematic illustration for the valence-space decoupling
by the IM-SRG evolution from $H(s=0) \to H(\infty)$ for two valence nucleons.
\label{fig:schematic_heff}}
\end{figure}

For open-shell nuclei, particle states $p$ can either be valence
particles or particle states above the valence space, which we denote
by $v$ and $q$ respectively. We want to decouple states that are not
in the valence space, spanned by $\ket{\Phi_v} = \{a_{v_1}^\dagger
\cdots a_{v_N}^\dagger\} \ket{\Phi_c}$, where $N$ is the number of
valence nucleons and $\ket{\Phi_c}$ is the reference state for the
core nucleus with $A_c$ nucleons. This can be realized by defining
$H^{\rm od}$ through the set of matrix elements
\begin{equation}
\{H^{\rm od}_1\} = \{f_{ph}, f_{qv}, \Gamma_{pp'hh'}, \Gamma_{pp'(vh
\, \text{or} \, hv)}, \Gamma_{(pq \, \text{or} \, qp)vv'}\} \,,
\label{eq:Hod1}
\end{equation}
where $p=v,q$, plus hermitian conjugates. As an alternative
generator choice $H^{\rm od}_2$, we also drive the one-body part 
to diagonal, so that we define
\begin{equation}
\{H^{\rm od}_2\} = \{H^{\rm od}_1, f_{pp'}, f_{hh'}\} \,.
\label{eq:Hod2}
\end{equation}
These generators both lead to a diagonal part $H_{1,2}^{\rm d}$ where
states outside the valence space are decoupled by the IM-SRG flow,
illustrated in Fig.~\ref{fig:schematic_heff}, leading to
\begin{equation}
P H_{1,2}^{\rm d}(\infty) Q = Q H_{1,2}^{\rm d}(\infty) P = 0 \,,
\end{equation}
with $P = \sum_{v} \ket{\Phi_v} \bra{\Phi_v}$ and $Q = 1-P$. The
off-diagonal parts in Eqs.~(\ref{eq:Hod1}) and~(\ref{eq:Hod2}) can
also be derived using the counting operator $C=\sum_{i} c_i
\{a_i^\dagger a_i\}$, with $c_i=1,0,-1$ for $q,v,h$ states,
respectively. The $C$ operator counts the number of excitations on top
of a valence-space state $\ket{\Phi_v}$. It is then straightforward to
verify that the above choices of $H_{1,2}^{\rm od}$ ensure $0= C
H(\infty) \ket{\Phi_v} = [C, H(\infty)] \ket{\Phi_v}$, which leads to
the decoupling of valence-space states from arbitrary excitations.

After the IM-SRG(2) evolution, the effective valence-shell Hamiltonian
is given by $H_{\rm eff} \equiv P H_{1,2}^{\rm d}(\infty) P-E_0^{A_c}$, 
where $E_0^{A_c}$ is the 0-body piece of the evolved Hamiltonian
corresponding to the ground-state energy of the core. We then solve a
reduced eigenvalue problem in the $N$ valence-particle space,
\begin{equation}
H_{\rm eff} \ket{\chi_n} = (E_n^{A} - E_0^{A_c}) \ket{\chi_n} \,.
\end{equation}

\begin{figure}[t]
\begin{center}
\includegraphics[width=0.48\textwidth,clip=true]{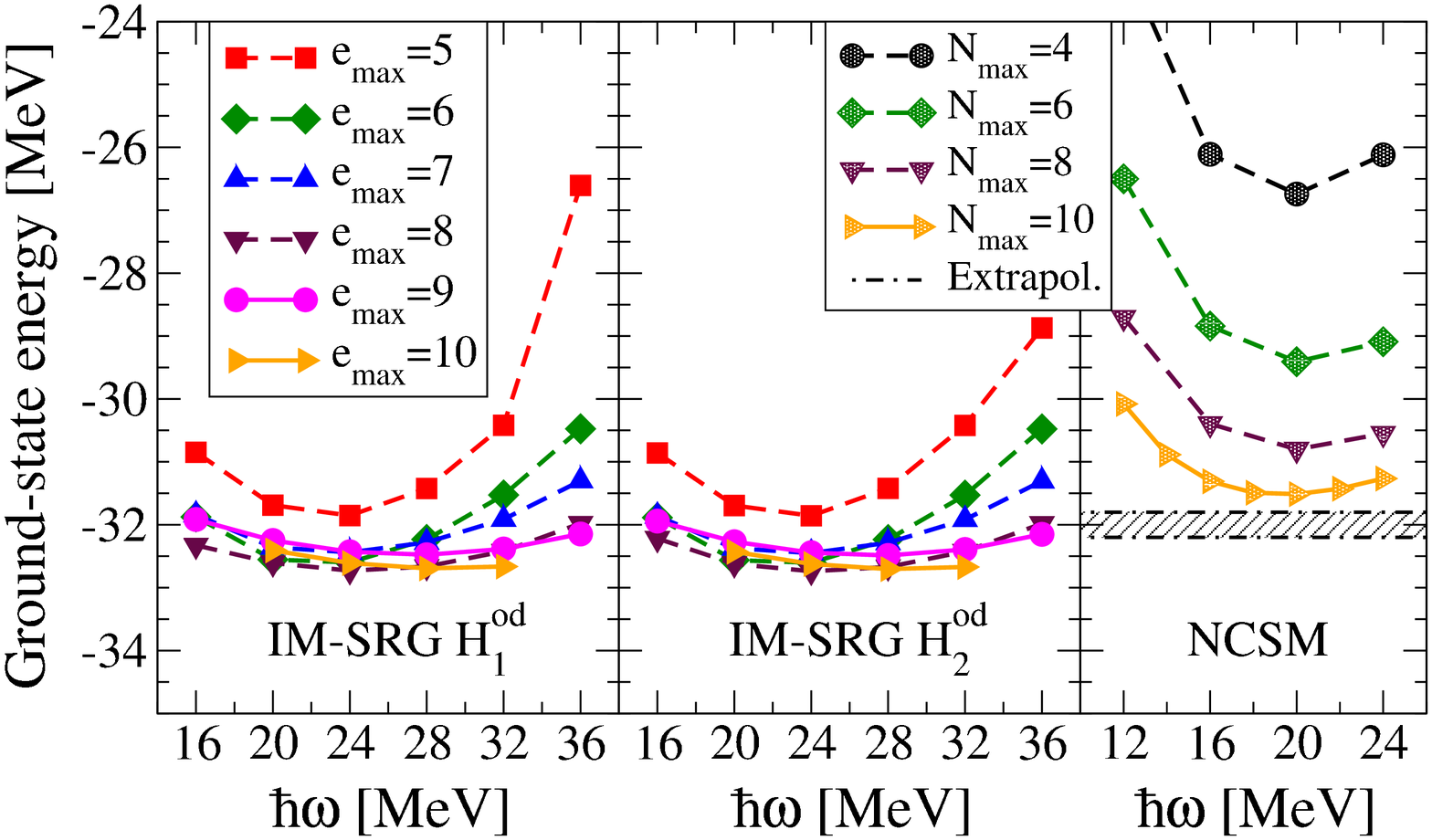}
\end{center}
\caption{The ground-state energy of $^{6}$Li versus
harmonic-oscillator parameter $\hbar \omega$ obtained by a
diagonalization of the IM-SRG(2) $H_{\rm eff}$ in the $p$-shell,
using generators $H_1^{\rm od}$ and $H_2^{\rm od}$ (left and center
panels). The flow equations are solved in the HF basis truncated to
$e_{\rm max}= \text{max}(2n+l)$ single-particle excitations. For
comparison we show in the right panel the convergence with
increasing $N_{\rm max}$ of the NCSM energy~\cite{Eric}, where the
dot-dashed band is the extrapolated result.\label{fig:valence_be}}
\end{figure}

{\it Results.--} We next present first applications of the IM-SRG to
two open-shell nuclei, $^6$Li and $^{18}$O, consisting of two valence
nucleons on top of the closed-shell nuclei $^{4}$He and $^{16}$O. All
results are based on the SRG-evolved ($\lambda = 2.0 \fmi$) N$^3$LO NN
potential of Ref.~\cite{N3LO}. We begin with $^6$Li, which is a
sufficiently light nucleus to allow a direct comparison of our
IM-SRG(2) results with exact diagonalizations of the 6-body problem
using the No-Core Shell Model (NCSM). For the IM-SRG calculations, the
flow equations are solved in the Hartree-Fock (HF) basis of the
initial Hamiltonian truncated to $e_{\rm max}={\rm max}(2n+l)$
single-particle excitations. The resulting $H_{\rm eff}$ is then
diagonalized in the $p$-shell to obtain the $^{6}$Li energy levels
with respect to the ground-state energy of the $^{4}$He core. There is
a subtlety that arises due to the self-bound nature of atomic
nuclei. As we wish to minimize spurious center-of-mass motion, we work
with the intrinsic Hamiltonian $H_{\rm int} = H - {\bf P}^2/(2mA)$
where ${\bf P} = \sum_i {\bf p}_i$ and $A=6$ for $^6$Li. Therefore,
the eigenvalues of $H_{\rm eff}$ correspond to the excitation energies
of $^6$Li with respect to the ground state of the unphysical $^4$He
nucleus obtained using $H_{\rm int}$ with $A=6$. Consequently, to get
the absolute ground-state energy of $^6$Li, we do a separate IM-SRG(2)
calculation of the ground state of the unphysical $^4$He core and add
this to the eigenvalues of $H_{\rm eff}$. We have checked that for
large $e_{\rm max}$ spaces, the center-of-mass factorizes as in
Ref.~\cite{cm}.

Figure~\ref{fig:valence_be} shows the convergence of the ground-state
energy of $^6$Li with increasing $e_{\rm max}$ excitations. The left
panels give the IM-SRG(2) results using the two different generators
$H_1^{\rm od}$ and $H_2^{\rm od}$, while the right panel shows NCSM
energies~\cite{Eric} for comparison. Since the single-particle $e_{\rm
max}$ truncation is different than the NCSM $N_{\rm max}$, the
convergence pattern of the two methods is expected to be different (an
$e_{\rm max}$ space for the same value is substantially larger). The
IM-SRG(2) ground-state energy converges to $-32.7(3) \mev$, where
contributions from normal-ordered three-body interactions are expected
to be repulsive (similar to triples correction in coupled-cluster
calculations)~\cite{Koshiroh}, in very good agreement with the
extrapolated NCSM value $-32.0(2) \mev$~\cite{Eric}. The generator
dependence in Fig.~\ref{fig:valence_be} is found to be very weak,
indicating that the error from truncating the IM-SRG equations to
two-body operators is indeed small.

\begin{figure}[t]
\begin{center}
\includegraphics[width=0.45\textwidth,clip=]{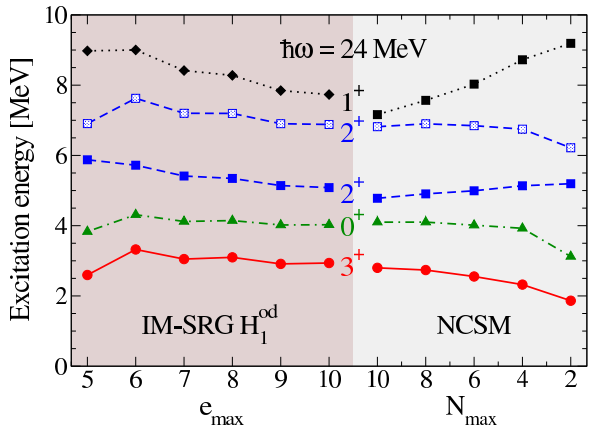}
\end{center}
\caption{Convergence as a function of $e_{\rm max}$ of the excitation
energies of $^{6}$Li obtained by diagonalizing the IM-SRG(2) $H_{\rm
eff}$ in the $p$-shell using the $H_1^{\rm od}$ generator. The HF
basis at a fixed $\hbar\omega=24 \mev$ is used for the IM-SRG(2)
calculations. For comparison we show the convergence with $N_{\rm
max}$ of the NCSM energies at the same $\hbar\omega$
value~\cite{Eric}.\label{fig:convergence_li6}}
\end{figure}

Next, we study the convergence properties of the low-lying excited
states of $^{6}$Li. The left panel of Fig.~\ref{fig:convergence_li6}
shows the convergence of the IM-SRG(2) spectrum as a function of
$e_{\rm max}$ at a fixed value of $\hbar\omega=24 \mev$. However, the
$\hbar\omega$ dependence is very weak in the HF basis for large
$e_{\rm max}$. The right panel shows the convergence of the NCSM
spectrum with $N_{\rm max}$. The low-lying $3^+$, $0^+$ and two $2^+$
states converge rather well and are in reasonable agreement with the
NCSM results. The high-lying $1^+$ state is not yet converged even at
the largest space for both methods. As the convergence is very poor in
the NCSM with a harmonic-oscillator basis, this could indicate that
this state has an extended structure.

Our results for $^6$Li are very encouraging and show that the IM-SRG
provides a new method to derive effective valence-shell Hamiltonians
that accurately reproduce the low-lying spectrum obtained with ab
initio methods, but at a polynomial scaling $\sim N_h^4 N_p^2$ with the
number of hole and particle orbits. Recently, other methods have been
explored for open-shell nuclei, including two-particle attached
coupled-cluster theory~\cite{2pattached}, which however leads to
nonhermitian effective Hamiltonians, and the NCSM with a
core~\cite{NCSMcore}, which requires a NCSM solution of the full
problem and is therefore limited to lighter nuclei.

\begin{figure*}[t]
\begin{center}
\includegraphics[width=0.4\textwidth,clip=]{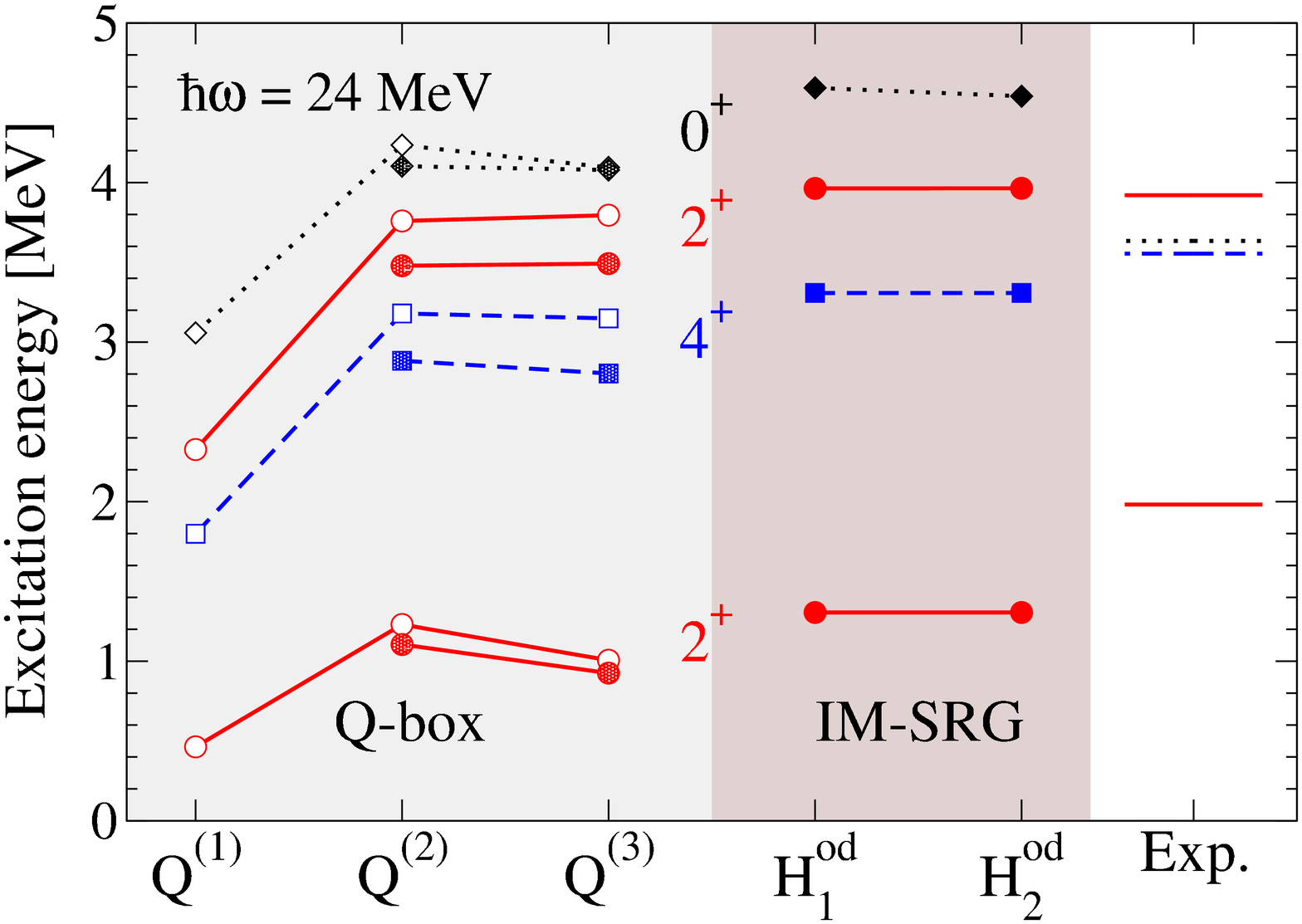}
\hspace*{10mm}
\raisebox{-1mm}{\includegraphics[width=0.4\textwidth,clip=]{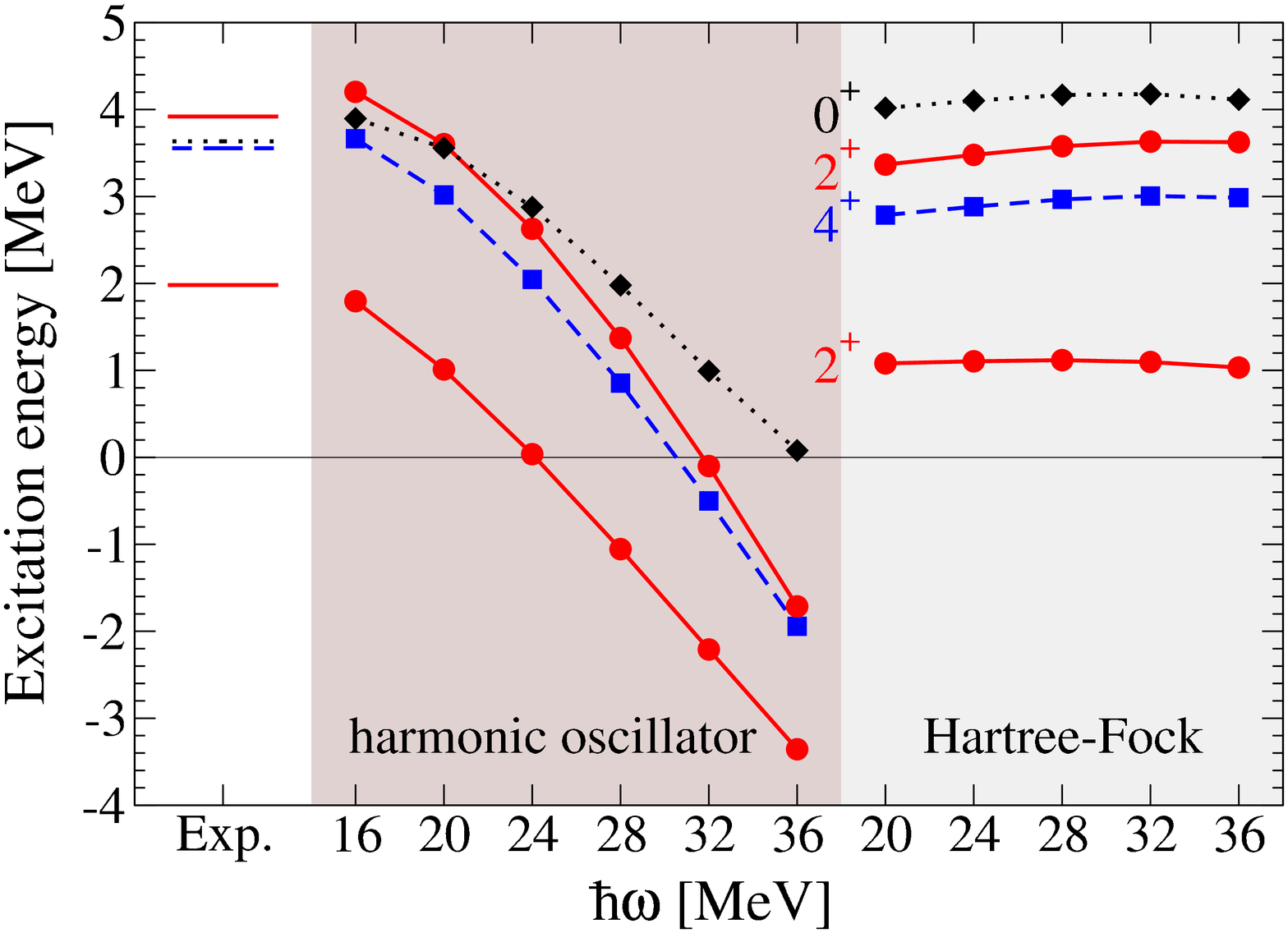}}
\end{center}
\caption{Left panel: Excitation energies of $^{18}$O obtained by
diagonalizing the IM-SRG(2) $H_{\rm eff}$ in the $sd$-shell,
compared with results obtained at first ($Q^{(1)}$), second
($Q^{(2)}$) and third ($Q^{(3)}$) order in the $Q$-box
expansion~\cite{Morten} for effective valence-shell interactions
(the filled symbols include higher-order folded-diagram
contributions). All results are for $\hbar\omega =24 \mev$, but a HF
basis is used. For comparison, we also show the experimental
energies~\cite{O18exp}. Right panel: Excitation energies of $^{18}$O
versus $\hbar\omega$ calculated at the second-order $Q$-box level
(plus folding) in a harmonic-oscillator and HF basis. For the
results of both panels, an $e_{\rm max}=8$ space was used, and to
simplify the comparison, the calculations used single-particle
energies from the USDb interaction~\cite{USD} for the
diagonalization in the $sd$-shell.\label{fig:o18}}
\end{figure*}

Turning to $^{18}$O, where an exact diagonalization of the 18-body
problem is out of reach, we compare our IM-SRG(2) results for the
spectrum in Fig.~\ref{fig:o18} to calculations based on diagrammatic
expansions (called the $Q$-box expansion) commonly used to derive
effective shell-model Hamiltonians~\cite{Morten}. In this context, one
can also understand our choices for $H_{1,2}^{\rm od}$ as follows. When
one derives effective interactions among valence nucleons using
perturbation theory, then the many-body diagrams contain at least one
vertex of $\Gamma_{pp'hh'}$, $\Gamma_{pp'(vh \, \text{or} \, hv)}$, or
$\Gamma_{(pq \, \text{or} \, qp)vv'}$. These interaction vertices are
precisely the off-diagonal part driven to zero under the IM-SRG
evolution. Therefore, the effective interactions among valence
nucleons are directly given by $P H_{1,2}^{\rm d}(\infty) P$ (only at
finite $s$, there would be perturbative corrections). For a
clear comparison to shell-model calculations for $^{18}$O, we
use the same empirical single-particle energies for the one-body part
of $H_{\rm eff}$ in both IM-SRG and $Q$-box calculations, that is we
replace the calculated one-body part in the IM-SRG by the empirical
USDb~\cite{USD} single-particle energies.

The left panel of Fig.~\ref{fig:o18} compares the low-lying $^{18}$O
excitation energies obtained by diagonalizing the $sd$-shell $H_{\rm
eff}$ derived from the IM-SRG(2) and the $Q$-box expansion. We also
give the experimental energies~\cite{O18exp}, although good agreement
with experiment is not required since three-nucleon (3N) forces are
not included in the initial Hamiltonian and we do not fine-tune the
single-particle basis to reproduce the experimental root-mean-square
radius (see below). All calculations are performed in the HF
basis. For the perturbative $Q$-box results, the open symbols
correspond to an $H_{\rm eff}$ that is calculated at first, second,
and third-order, while the solid symbols include higher-order
folded-diagram contributions to remove the energy dependence of
induced interaction vertices~\cite{Morten}. For the IM-SRG(2) results,
as for $^6$Li, we observe negligible differences in the calculated
spectra for the two generators $H_1^{\rm od}$ and $H_2^{\rm od}$. This
implies that the truncation of the flow equations to two-body
operators is a very good approximation. The IM-SRG(2) energies are
similar to the results based on the perturbative $Q$-box expansion,
where some differences from the ``best'' results ($Q^{(3)}$ plus
folded-diagram contributions) are expected because the IM-SRG is a
nonperturbative method that includes many higher-order terms.

The poor agreement with experiment in Fig.~\ref{fig:o18} compared to
conventional shell-model calculations might be surprising. The reason
for this discrepancy is that conventional calculations include
additional phenomenology that improves agreement with experiment, but
weakens the connection with the underlying Hamiltonian and microscopic
many-body theory. This can be understood from the right panel of
Fig.~\ref{fig:o18}, which displays the $\hbar\omega$ dependence of the
$^{18}$O excitation energies for second-order $Q$-box calculations
performed in the harmonic-oscillator (HO) and the HF basis. The
HO-based spectrum exhibits a very strong $\hbar\omega$ dependence,
while calculations in the HF basis are nearly independent of
$\hbar\omega$. Conventional calculations of $H_{\rm eff}$ work in the
HO basis with $\hbar\omega \approx 45 A^{-1/3} - 25 A^{-2/3}$ chosen
to give the same root-mean-square radius as a sphere of uniform
density. The fine-tuning of $\hbar\omega$ can therefore be understood
as a phenomenological means to build in the correct saturation
properties of nuclei, which are known to be deficient in ab-initio
calculations starting from Hamiltonians without three-nucleon
forces~\cite{PPNP}. Our IM-SRG calculations present a
microscopically-derived $H_{\rm eff}$, but the incorrect saturation
properties of the initial NN-only Hamiltonian translate into a poor
description of the $^{18}$O spectrum compared to empirical
calculations carried out in a HO basis (at $\hbar\omega \approx 14
\mev$ for $^{18}$O).

{\it Conclusions.--} We have shown that the IM-SRG can be successfully
generalized to open-shell systems and to a nonperturbative derivation
of effective valence-shell Hamiltonians. The IM-SRG evolution
decouples the physics of valence nucleons from the full Hilbert space,
enabling exact diagonalizations in the valence space that are
impossible in the full problem where all nucleons are active. First
results were presented for $^6$Li, with ground-state and excited-state
energies in very good agreement with ab-initio methods. We then
applied the IM-SRG to $^{18}$O and compared our results to those
obtained from conventional perturbative calculations of $H_{\rm eff}$,
demonstrating that the IM-SRG provides a first viable nonperturbative
approach to derive effective interactions for the shell model from
nuclear forces. Work is in progress to extend the IM-SRG to extended
valence spaces, to effective operators and to the inclusion of
three-nucleon forces.

\begin{acknowledgments}
We thank H.\ Hergert and T.\ Otsuka for useful discussions.
This work was supported in part by the HPCI, the UNEDF SciDAC
Collaboration under DOE Grant DE-FC02-07ER41457, the NSF under Grant 
No.~PHY-1068648, the Helmholtz Alliance Program of the Helmholtz
Association, contract HA216/EMMI ``Extremes of Density and
Temperature: Cosmic Matter in the Laboratory'', and the DFG through
grant SFB 634.
\end{acknowledgments}

\end{document}